# Multi Spectral Switchable Infra-Red Reflectance Resonances in Highly Subwavelength Partially Oxidized Vanadium Thin Films

Ashok P, Yogesh Singh Chauhan, and Amit Verma

Department of Electrical Engineering, Indian Institute of Technology Kanpur, Kanpur 208016, India

Email: ashok@iitk.ac.in and amitkver@iitk.ac.in

ABSTRACT: Phase transition materials are promising for realization of switchable optics. In this work, we show reflectance resonances in the near-infrared and long-wave infrared wavelengths in highly subwavelength partially oxidized Vanadium thin films. These partially oxidized films consist of a multilayer of Vanadium dioxide and Vanadium as shown using Raman spectroscopy and four-probe measurements. As Vanadium dioxide is a phase transition material that shows insulator to metal phase transition at 68 °C, the observed infra-red resonances can be switched with temperature into a high-reflectance state. The wavelength of these resonances are passively tunable as a function of the oxidation duration. The obtained reflectance resonance at near-infrared wavelength red shifts from 1.78 µm to 2.68 µm with increasing oxidation duration while the long-wavelength infrared resonance blue shifts from 12.68 µm to 9.96 µm. To find the origin of the reflectance resonances, we model the reflectance spectra as a function of the oxidation duration using the transfer matrix method. The presented model captures the dual reflectance resonances reasonably well. These passive wavelength- tunable and switchable resonances with easy to fabricate lithography-free multilayer structure will be useful for multispectral applications such as camouflage, spectral selective microbolometer, and thermal management.

1. Introduction

Smart control of light waves is attractive for next-generation optics. Especially, wavelength-selective and switchable/reconfigurable resonances are useful for applications such as spectral selective microbolometer [1-3], multispectral camouflage [4], hyperspectral imaging [5], modulators, optical switches, and memories [6]. Typically metamaterials are used for obtaining wavelength-selective absorbers/resonances. For example, a metamaterial may consist of a trilayer structure of a patterned metal resonator, dielectric, and bottom conducting plane. By varying the size of the metal resonator or varying the dielectric thickness of the metamaterial [1,7], wavelength-selective resonances can be achieved. Multi or dual-band resonance in metamaterials have been demonstrated in THz [8] and Mid-wave Infrared wavelength (MWIR) [9]. However, fabricating these metamaterials to obtain single or multi-band resonances need multiple deposition and complex patterning/lithography processes.

Vanadium dioxide ($VO_2$) is a natural metamaterial that shows a large change in refractive index, particularly in the infrared wavelengths, while transitioning from insulator to metal phase at 68 °C [10]. Incorporating $VO_2$ thin film in optics offers both switchable and wavelength-tunable resonances without pattering. Near perfect absorption/resonance is achieved in Long-wave Infrared (LWIR) wavelength using ultra-thin $VO_2$ film on the sapphire substrate [10] and in Mid-Wave Infrared (MWIR) wavelength using $VO_2$ film on ITO substrate [11]. This resonance is similar to asymmetric Fabry-Perot resonance, but the thickness of the $VO_2$ film is much smaller than the wavelength of light. These wavelength-selective resonances are single band resonance either in MWIR or LWIR wavelength and can be tuned by varying the thickness of the $VO_2$ layer or tailoring the plasma frequency of the substrate [12]. Dual switchable resonances have been demonstrated by integrating $VO_2$ with periodic pattern of micro-disc shaped gold metal [13]. Multiple resonances have also been shown using the trilayer structure of $Ge/VO_2/Al_2O_3$. These multiple resonances were however Fabry Perot resonances mainly from the 1 micron thick Ge layer which is comparable to the wavelength of light [14].

In this work, we demonstrate two reflectance resonances, one in Near-Infrared (NIR) and another in LWIR wavelengths, which can be obtained in a single step by partially oxidizing a highly subwavelength Vanadium thin film in air. The wavelength of these resonances are passively tunable by changing the oxidation duration and reflectance is switchable because of the presence of $VO_2$ layer in the top part of the structure. We also present a $VO_2$/V/Sapphire multilayer model which reasonably explains the observed resonances and reflectance switching in the partially oxidized V thin films.

2.  **Experimental Procedure**

We used RF sputtering to deposit vanadium thin film on a c-plane sapphire substrate. During the deposition, we maintain a chamber pressure of 2 Pa by introducing argon gas (purity of 99.999%) into the chamber. The sputtering was done from a 2-inch vanadium target, and the sapphire substrate was maintained at 450 ˚C during the deposition. Thickness of the Vanadium film was 130 nm as measured using a KLA-Tencor stylus profilometer. After the deposition, the sample was diced into multiple pieces and subsequently oxidized for different oxidation durations $t_{oxd}$ = 2 min-8 min, in open atmosphere on a hot plate at 450 ˚C followed with quenching using a cold plate at room temperature. More details of the oxidation process can be found in our previous works [15-17]. To measure the temperature-dependent optical reflectance of all the oxidized samples, we used an IR microscope (Agilent, Cary 600) connected to a FTIR spectrometer (Agilent, Cary 660). Raman spectroscopy (using an excitation laser of wavelength 532 nm) and temperature-dependent four-probe electrical resistance measurements were also used to characterize all the oxidized samples.

3.  **Results and Discussion**

Fig.1(a) and 1(b) show FTIR reflectance measured at room temperature (27 ˚C) and high temperature (100 ˚C), respectively, for Vanadium film and oxidized samples. The unoxidized vanadium thin film deposited at 450 ˚C shows high reflectance with no resonances in the infrared wavelength. However, as oxidation happens, a single resonance appears in NIR wavelength for 2 min oxidized sample. As oxidation time increases, another resonance starts to appear at LWIR. Films oxidized for 3-7 min show strong dual resonances at NIR and LWIR wavelengths. After 7 min of oxidation, the film shows single resonance at LWIR wavelength. High-temperature reflectance measurement (Fig. 1(b)) shows monotonically increasing reflectance with increasing oxidation time and no resonances in the infrared wavelength. Fig.1(c) shows the reflectance resonance wavelengths at NIR and LWIR for all the oxidized samples. With an increase in oxidation time, resonance at NIR wavelength gets red-shifted from 1.78 µm to 2.68 µm, while resonance at LWIR gets blue-shifted from 12.68 µm to 9.96 µm. Fig.1(d) shows reflectance switching at NIR/LWIR and peak reflectance switching of all the samples. We observe that reflectance switching at 10 µm increases monotonically with oxidation time and peaks at $t_{oxd}$ = 8 min. In contrast, reflectance switching at 1.8 µm continuously decreases as a function of increasing oxidation time. The peak switching of more than 50 % was observed for all the oxidized samples. Peak switching of 70 % was observed for $t_{oxd}$ = 7 and 8 min samples.

To find the phases present in the oxidized films, we characterized the samples using Raman spectroscopy and temperature-dependent four-probe measurements. Atmospheric thermal oxidation of Vanadium thin films has three sequential regimes. In the first regime, initially, $VO_2$ forms at the surface of Vanadium film. As the oxidation time increases, the $VO_2$ content of the film increases, and finally, the vanadium film is completely oxidized to form $VO_2$ thin film. In the second regime, $V_2O_5$ formation happens, which leads to a mixed phase of $VO_2$ and $V_2O_5$. In the third regime, only phase pure $V_2O_5$ forms for a longer oxidation duration. These oxidation regimes are experimentally observed in previous works [15, 17-19].

Fig.2(a) shows Raman spectra of all the oxidized samples. All samples show only $VO_2$ Raman peaks except the $t_{oxd}$ = 8 min film which shows both $VO_2$ peaks and c-plane sapphire substrate Raman peak [20]. The films oxidized for $t_{oxd}$ < 8 min possibly have unoxidized metallic vanadium film, which blocks substrate Raman peaks, while $t_{oxd}$ = 8 min film is completely oxidized, and the absence of metallic vanadium film shows substrate peak. Raman spectroscopy results suggest that the thermal oxidation of Vanadium is clearly in the first regime where only $VO_2$ forms. Fig.2(b) shows the thermal oxidation growth model of Vanadium in the first regime where $VO_2$ forms at the surface of Vanadium thin film. As oxidation continues, the thickness of the $VO_2$ layer increases, and finally, Vanadium film completely oxidizes into $VO_2$.

To confirm these results, we measured the temperature-dependent resistance of the samples using four-point probe measurements. Fig.2(c) shows four-probe resistance measured at 40 ˚C and 110 ˚C as a function of the oxidation duration. All oxidized samples show low resistance at 40 ˚C, except $t_{oxd}$ = 8 min sample, which shows high resistance of more than two orders compared to other films. This resistance trend is mainly due to the unoxidized vanadium film below the $VO_2$ layer, which shunts the electric current in the films up to $t_{oxd}$ = 7 min. The film oxidized for $t_{oxd}$ = 8 min lacks parallel metallic vanadium conduction path, which increases the measured resistance in the film. $t_{oxd}$ = 8 min film also shows reversible resistance switching (Fig.2(d)) of more than two orders of magnitude. Thus, four-probe resistance measurements also confirm that thermal oxidation of Vanadium follows the first regime for the $t_{oxd}$ durations in this study.

To understand the origin of reflectance resonances and switching modulation in the films as a function of the oxidation duration, we model the reflectance spectra using the transfer matrix method [21]. We extract the complex refractive index of the deposited Vanadium using the Drude-Lorentz oscillator model to match the reflectance spectra of $t_{oxd}$ = 0 min unoxidized film. Similarly, we fitted the reflectance spectra of $t_{oxd}$= 8 min for extracting the complex refractive index of $VO_2$. According to the Drude-Lorentz oscillator model [11], the complex dielectric response of any material at frequency ω is given as:

$$\varepsilon(\omega) = \varepsilon_\infty - \frac{\omega_p^2}{\omega(\omega+i\omega_c)} + \sum_j \frac{f_j}{1-\frac{\omega^2}{\omega_j^2}-i\gamma_j\omega/\omega_j} \qquad (1)$$

where, $\varepsilon_\infty$ is the high frequency permittivity, $\omega_p$ corresponds to plasma frequency and $\omega_c$ is the collision frequency. The last part is the Lorentz oscillator, where $\omega_j$ is the resonance frequency, $f_j$ is the strength of oscillator, and $\gamma_j$ is the line width. The extracted model parameters of Vanadium and $VO_2$ are summarized in Table 1.

Using these refractive indices, we calculated the reflectance of Air/VO$_2$/V/Sapphire multilayer using the transfer matrix method. For $t_{oxd}$ = 0 min sample, V thickness of 130 nm was used with no VO$_2$ while for $t_{oxd}$ = 8 min sample, VO$_2$ thickness of 260 nm was used with no V. We used double thickness for VO$_2$ compared to starting V because that is the expansion factor from V to VO$_2$ due to oxidation [22]. For intermediate $t_{oxd}$ between 0 min and 8 min, V is partially oxidized to VO$_2$. Assuming a linear oxidation law, we calculated the VO$_2$ thickness (double the thickness of oxidized V) and remaining V thickness to calculate the reflectance of resulting Air/VO$_2$/V/Sapphire stack. Fig.3 shows the calculated reflectance spectra at room temperature as a function of the oxidation duration. The transfer matrix model is able to capture the two resonances (reflectance minima) at room temperature in the NIR and LWIR wavelengths. The experimental resonance positions are also added on the figure for comparison. The model captures the resonance red-shift at NIR and resonance blue-shift at LWIR with an increase in the oxidation duration. These resonances shift mainly from the thickness variation of Vanadium and VO$_2$ film as a function of the oxidation duration. The model, however, does not capture the precise NIR resonance position for higher oxidation times ($t_{oxd}$ > 4 min). The discrepancy might be due to variation in the refractive indexes of Vanadium and Vanadium dioxide (currently extracted from $t_{oxd}$ = 0 min and $t_{oxd}$ = 8 min samples, respectively) with the oxidation duration. Further study is needed to understand this difference.

It is worth mentioning that the sapphire substrate is opaque in the LWIR (10-15 µm) [10] range and Vanadium metal is almost opaque in the NIR range, so the absorptivity of the VO$_2$/V/Sapphire stack is A= (1-R), where R is reflectivity in the infrared spectrum. Thus, the reflectance resonance at NIR and LWIR is due to strong absorption in the structure. Moreover, as thermal emission is proportional to absorption via Kirchhoff's law [23], demonstrated structure can also provide tunable emission as a function of temperature in the NIR and LWIR. VO$_2$ has the advantage that an electric field can also trigger the phase transition at an ultrafast nanosecond timescale [24, 25]. Combining the VO$_2$/V/sapphire structure with electric field triggering can enable ultrafast multispectral tuning of the reflectance/emissivity. The VO$_2$/V/Sapphire stack will therefore be useful for multispectral absorption (bolometer), emission (thermal emitters) and active infrared camouflage applications.

## 4. Conclusion

In conclusion, we used partial thermal oxidation of vanadium thin films to demonstrate passively tunable reflectance resonances and reflectance switching in NIR and LWIR wavelengths. In addition, we have modeled the reflectance spectra as a function of the oxidation duration. The model captures and explains the origin of reflectance resonance shift with the increasing oxidation duration. This demonstration of wavelength-selective reflectance resonances and multi-spectral reflectance switching using simple thermal oxidation with no complex patterning steps will enhance the optical applications such as multi-spectral camouflage and radiators based on phase transition oxide materials.

**Acknowledgements**

This project was supported by Science and Engineering Research Board, India via Imprint Grant No. IMP/2018/000404.

Table 1. The optimized Lorentz- Drude oscillator parameters of Vanadium (Thickness 130 nm) film and Vanadium dioxide (Thickness 260 nm) in both the insulating and the metallic phase.

| | $\varepsilon_\infty$ | $\hbar\omega_p$ (eV) | $\hbar\omega_c$ (eV) | $\hbar\omega_1$ (eV) | $\hbar\omega_2$ (eV) | $\hbar\omega_3$ (eV) | $\gamma_1$ | $\gamma_2$ | $\gamma_3$ | $f_1$ | $f_2$ | $f_3$ |
|---|---|---|---|---|---|---|---|---|---|---|---|---|
| **Vanadium** | 2.526 | 6.624 | 8.924 | 0.188 | 0.040 | 0.086 | 0.397 | 3.783 | 0.12 | 68 | 700 | 2.1 |
| **VO$_2$ (Insulating)** | 11.56 | - | - | 0.062 | - | - | 0.022 | - | - | 18.07 | - | - |
| **VO$_2$ (Metallic)** | 25.94 | 5 | 0.376 | - | - | - | - | - | - | - | - | - |

**FIGURES**

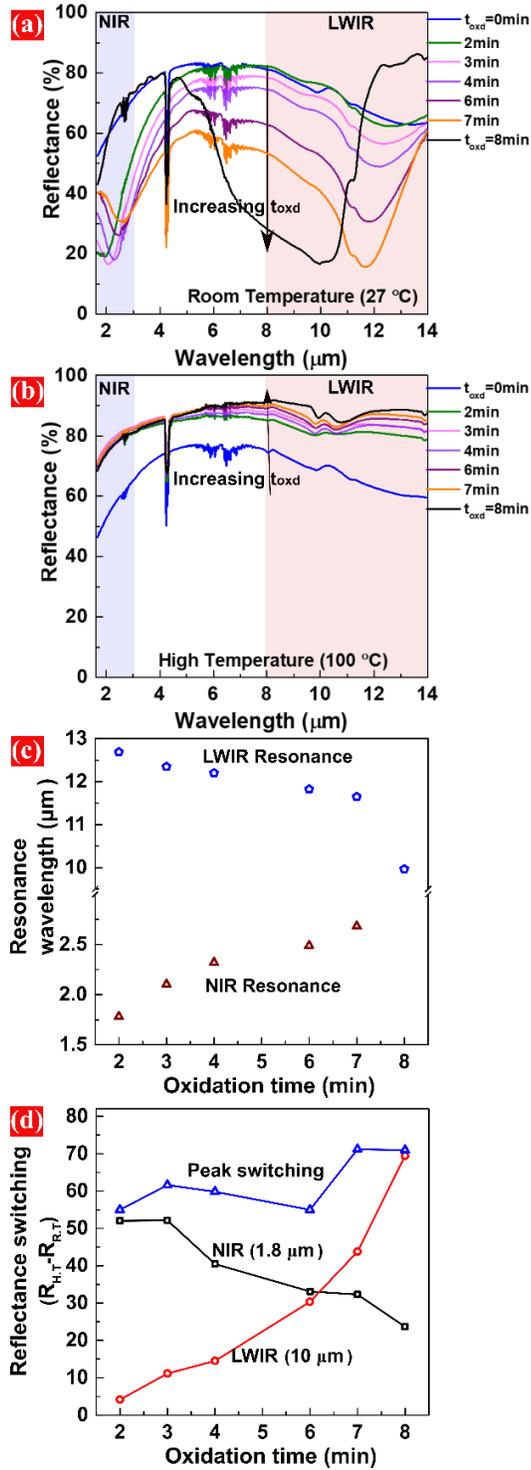

FIG. 1 FTIR reflectance spectra measured at (a) room temperature (27 ˚C) and (b) high temperature (100 ˚C) for all the samples.(c,d) Resonance shift and Reflectance switching as a function of oxidation duration for the oxidized samples (line is guide to the eye in (d)).
.

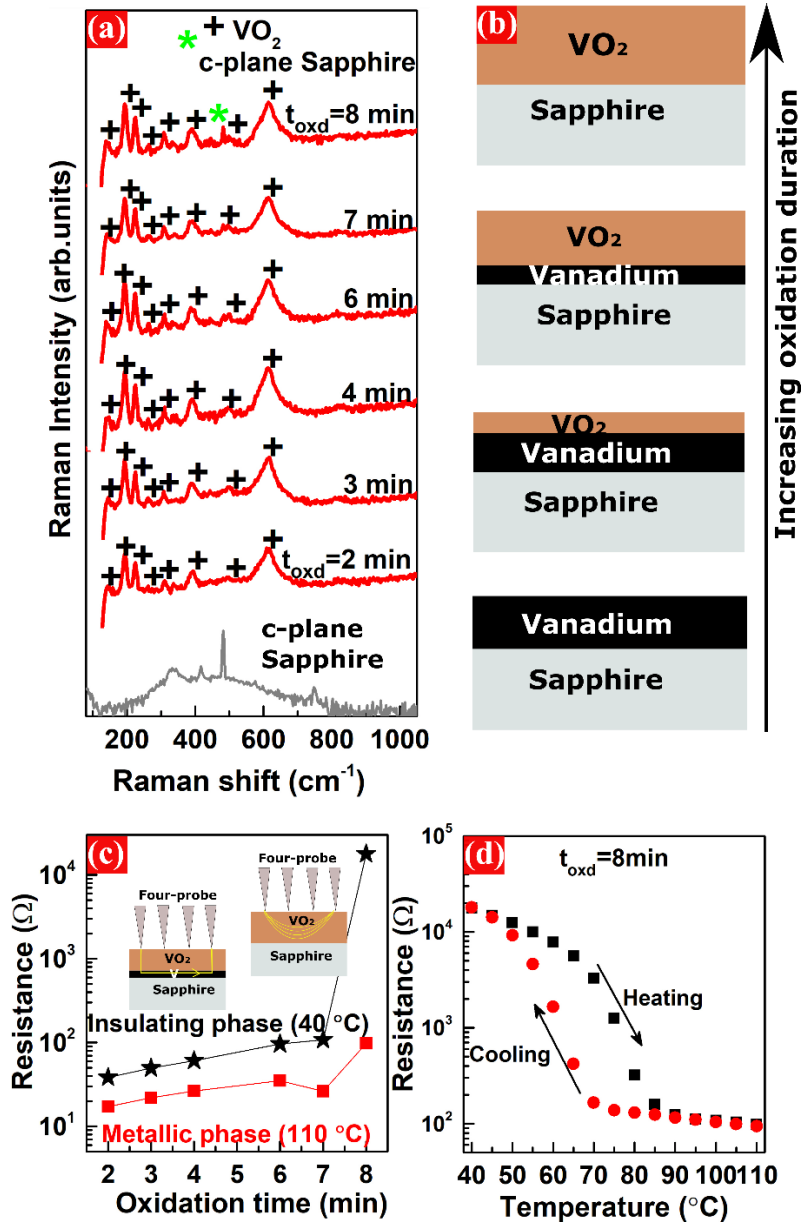

FIG. 2. (a) Raman spectra of all oxidized Vanadium films. (b) Oxidation growth model of Vanadium thin film. (c) Four-probe resistance measured at 40 °C and 110 °C for all the oxidized samples (inset shows current path for oxidized samples in the presence and absence of V). (d) Reversible resistance switching of $t_{oxd}$= 8 min $VO_2$ sample.

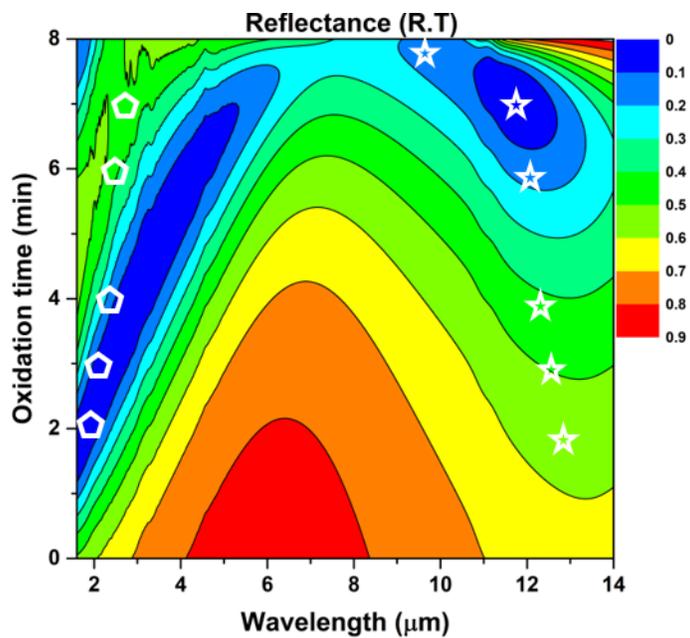

FIG. 3. Calculated reflectance spectra at room temperature as a function of the oxidation duration. Symbols reperesent the measured resonances at NIR( pentagon) and LWIR (star).(The intensity of reflectance is shown by color code).